\documentclass[twocolumn,showpacs,preprintnumbers,amsmath,amssymb,aps,prl]{revtex4}

\usepackage{graphicx}% Include figure files

\begin{document}

\title{A simple algorithm based on fluctuations to play the market}
% Force line breaks with \\

\author{\bf L. Gil}
\affiliation {Institut Non Lin\'{e}aire de Nice, UMR 6618 CNRS-UNSA, 1361
Route des lucioles, 06560 Valbonne - Sophia Antipolis
\\ FRANCE}

\date{\today}

\begin{abstract}
In Biology, all motor enzymes operate on the same principle: they trap favourable brownian fluctuations in order to generate directed forces and to move. Whether it is possible or not to copy one such strategy to play the market was the starting point of our investigations. We found the answer is yes. In this paper we describe one such strategy and appraise its performance with historical data from the European Monetary System (EMS), the US Dow Jones, the german Dax and the french Cac40.
\end{abstract}


\maketitle
\section{Introduction}
\label{intro}
In the 1970s, the Efficient Market Hypothesis (EMH) assumed that the actions of all traders and external news result collectively into so large fluctuations that any predictability pattern is neither detectable nor exploitable \cite{Bachelier, Fama70}. No predictions can be performed from the knowledge of the past prices and therefore to make profit, one has to be informed (possibly illegaly) of the future evolutions of currencies or securities. 

Now, even \cite{PBall} the champions of EMH are forced to recognize that first, several disagreements between the efficient market theory and genuine experimental observations have been exhibited \cite{ExcessVolatility,Thaler2005,RandomWalk,BehavioralFinance}, and second, that the central assumption that every trader or company acts in the same self-interested way rational, cool and omniscient is only a gross caricature, in strong contradiction with the experimental psychological observations \cite{BoundedRationality}. The EMH hypothesis must be viewed as a first order (and nevertheless useful) approximation describing an average idealized behavior.

Since financial markets are not in strictly perfect equilibrium, it should be possible to beat the market using algorithms based on fluctuations. In this paper we exhibit and describe one such an algorithm and appraise its performance with the german mark(DEM), the english pound (GBP) and french franc (FRF) currencies during the European Monetary System (EMS) from 1980 to 1992, and with the securities which compose the US Dow Jones, the german Dax and the french Cac40 between 2000 and 2006. The results are quite impressive, especially with the EMS, where average costless yearly return up to $50\%$ are obtained.

\section{Building of the algorithm}
\label{building}
The present algorithm is sustained by two basic ideas: first, renounce to predict the future prices, second, take full advantage of the stochastic characteristic of their temporal evolution (we will show that the absence of fluctuations leads to vanishing returns). The algorithm takes its origin from an analogy with the mechanism called noise induced transport, putted into evidence in Physics and which has been found to be deeply involved in Biology.

In Physics, the Maxwell's demon as well as the Feynman's ratchet \cite{FeynmanRatchet} are very famous illustrations of one of the main implication of the second law of thermodynamics, i.e. that useful work cannot be extracted from equilibrium fluctuations.  In 1990s, it has been realized that this restriction does not hold anymore for out of equilibrium situations and a lot of theoretical \cite{Magnasco93,Bier94,Peliti94}, as well as numerical and experimental \cite{Prost94} studies, have reported onto the possibility for asymmetric potentials to rectify non colored fluctuations and to lead to a coherent response from unbiased forcing. The paradigm of such situation is the flashing brownian ratchet. In this system, brownian particles are subjected to a spatial periodic potential, which breaks the parity symmetry and is periodically turned on and off with time. When off, the particles symmetrically spread. When on, parity symmetry is broken. As a result, a noise induced transport takes place. Numerous applications of these ideas have been found in Biology where brownian fluctuations are ubiquitous and unimaginably tumultuous \cite{Prost94,Oster2002}. For example, the proteins (kinesin and dynein) which are involved in the transport between the nucleus and the membrane of a cell, are just allowed to attach and detach themselves from the periodic chiral pilings up of moleculs called microtubules. However they move!

In the field of finance, spatial position of the protein stands for the security price, transport means positive return, brownian fluctuations stand for random walk of the price returns and out of equilibrium situations correspond to non efficient market. Stationary attached position correspond to short situation while detached (and then sensitive to fluctuations) to long one. The old maxim "buy at low price and sell at hight" will play the role of the asymmetry of the potential. Of course, at this stage, we are still left with the crucial point to precisely define what "low" and 'hight" means (we will do it in the following). But what has to be learned from the biological analogy, is there is no need of a (intelligent) human decision, no need of a "deep" and "smart" understanding of the market, and that a systematic, repetitive and programmable trading is enought. After all, proteins are not clever!

Now we built up step by step the algorithm base on fluctuations we will use to play the market. As a starting point, we restrict ourselves to a market composed of only one security, whose price is $X_{1}(t)$. Time is discret and the interval between two successive times is constant and arbitrary set to one. We first assume that 
\begin{equation}
X_{1}(t_{n})=A_{1}+a_{1} \phi_{1}(t_{n})
\label{basic}
\end{equation}
where $\phi_{1}$ is an independent aleatory variable equal to $\pm 1$ with probability ${{1}\over{2}}$. There are no longscale trends because $A_{1}$ is constant with respect to time, only binomial fluctuations of amplitude $a_{1}$ with $a_{1}<A_{1}$.  It is obvious that, for such a system, the average return of the buy and hold strategy is just $0$. By analogy with motor enzymes which trap favourable brownian fluctuations in order to progress, the aim of our Maxwell's demon will be to select the fluctuations of price which correspond to an increase of security number. His behavior (MD1) is defined as
\begin{enumerate}
\item The demon is playing once at each time $t_{n}$. 
\item At each time $t_{n}$, the demon is in possession of either securities or money, but neither a mixing of both.
\item At time $t_{n}$, the choice between security and money is ruled by

\noindent
{\bf R1:} If $X_{1}(t_{n})>A_{1}$ then the demon tries to get money. Therefore, if he was in possession of securities, then he sells them.

\noindent
{\bf R2:} If $X_{1}(t_{n})<A_{1}$ then he tries to get securities. Therefore, if he was not in possession of securities, then he buys them.
\end{enumerate}

\begin{figure}[!h]
{\resizebox*{8.5cm}{6.0cm}{\includegraphics{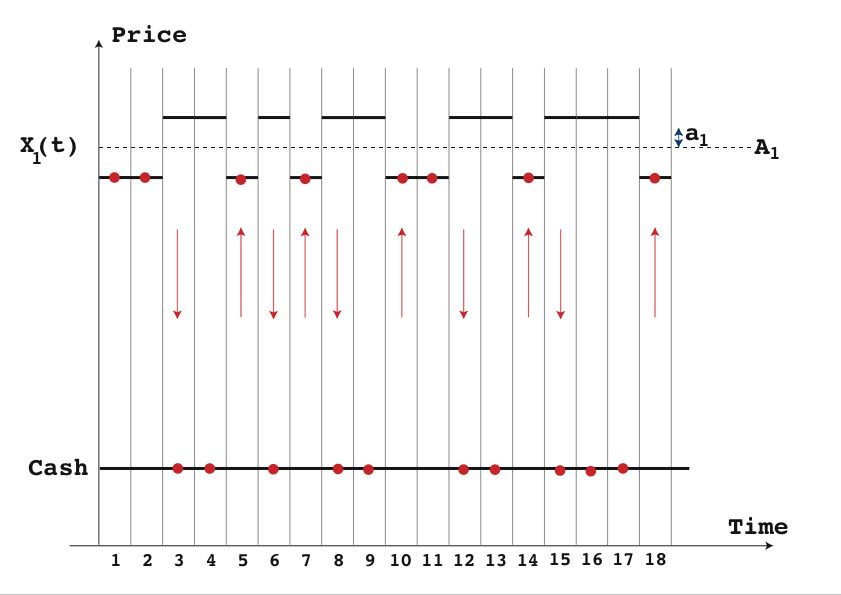}}}
\caption{\label{fig1} A typical trajectory showing the choice (black disk) between security ($X_{1}$) and money versus time. The downward arrows correspond to $(R_{1})$ case while the upward ones to $(R_{2})$. The lack of arrows correspond to situations where the choice at $t_{n}$ already corresponds to the position at $t_{n-1}$.}
\end{figure}

\begin{figure}[!h]
{\resizebox*{8.6cm}{5.0cm}{\includegraphics{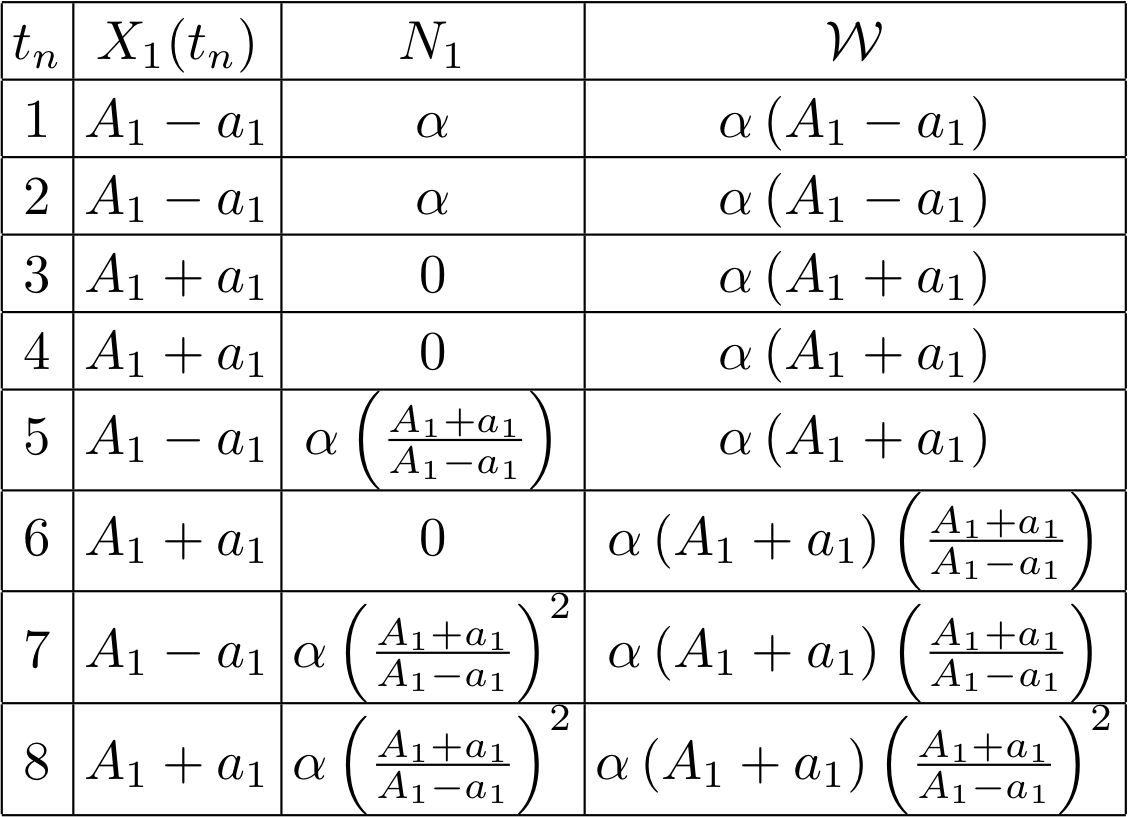}}}
\caption{\label{fig2} Evolution of the number of security ($N_{1}$) and of the total wealth ${\cal W}$ for the given price trajectory displayed in fig.\ref{fig1}. Althought the time $t$$=$$3$, $t$$=$$6$ and $t$$=$$8$ configurations are strictly identical, the wealth has been multiplied respectively by $\delta$, $\delta^2$ and $\delta^3$.}
\end{figure}

Fig.\ref{fig1} shows both a typical price trajectory $\{$$X_{1}(t_{1})$, $X_{1}(t_{2})$.... $X_{1}(t_{n})$$\}$ and the associated  demon's behavior while fig.\ref{fig2} describes the corresponding time evolution of his wealth ${\cal W}$ and the number $N_{1}$ of securities which are in his possession. Despite the dynamic is limited to the 8 first days, one can already observe that, although the third and sixth day price configurations are the same, the wealth has been increased by a multiplicative factor 
$\delta$$=$${{A_{1}+a_{1}}\over{A_{1}-a_{1}}}$
greater than 1. Defining the daily return $r(t_{k})$ and the cumulative daily return $R(t_{n})$ as
\begin{equation}
r(t_{k})=log\left({{{\cal W}(t_{k})}\over{{\cal W}(t_{k-1})}}\right)
\qquad
R(t_{n})=log\left({{{\cal W}(t_{n})}\over{{\cal W}(t_{1})}}\right)
\label{return}
\end{equation}
and averaging over the $2^{n}$ possible trajectories, we then easily obtain 
$\left< R(t_{n}) \right>={{n}\over{8}} log(\delta) $.

\begin{figure}[!h]
{\resizebox*{8.5cm}{6.0cm}{\includegraphics{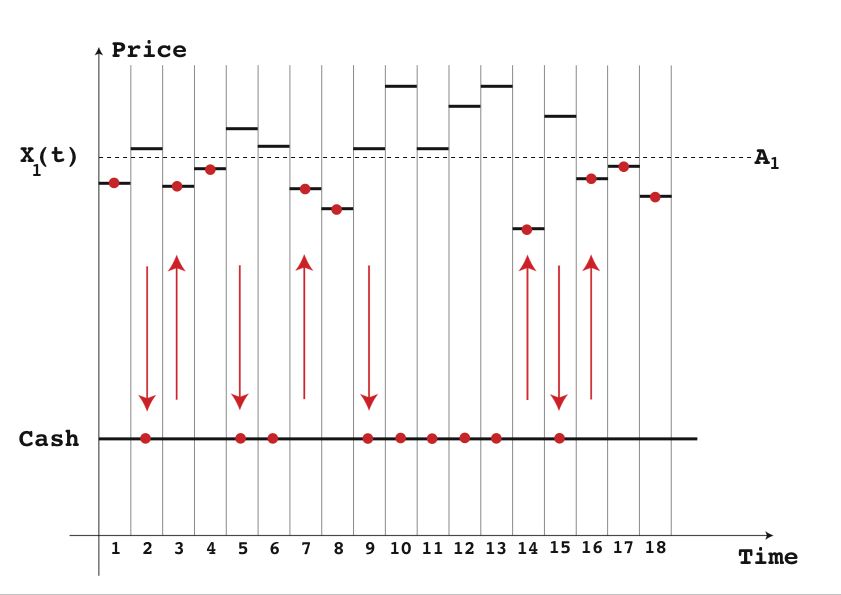}}}
\caption{\label{fig3} A typical trajectory showing the selected security (black disks) versus time for an arbitrary $\phi$ distribution.}
\end{figure}

\begin{figure}[!h]
\includegraphics[width=8.7cm]{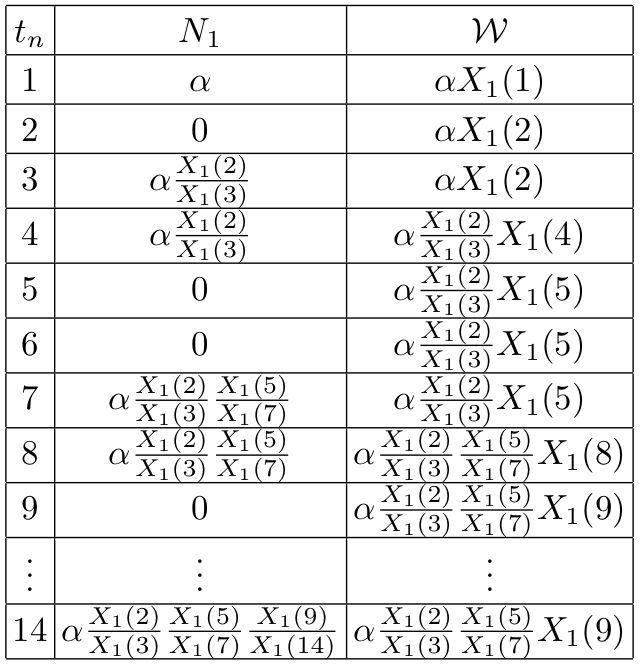}
\caption{\label{fig4}Evolution of the number of security ($N_{1}$) and of the total wealth ${\cal W}$ for the 14 first days corresponding to the trajectory displayed in fig.\ref{fig3}. It is crucial to note the increasing of the non vanishing $N_{1}$ numbers. For example, because ${{X_{1}(t_{2})}\over{X_{1}(t_{3})}}$, ${{X_{1}(t_{5})}\over{X_{1}(t_{7})}}$ and ${{X_{1}(t_{9})}\over{X_{1}(t_{14})}}$ are all $>1$, we are sure that $N_{1}(t_{14})>N_{1}(t_{7})>N_{1}(t_{3})>N_{1}(t_{1})$.}
\end{figure}

\begin{figure}[!h]
{\resizebox*{8.5cm}{6.0cm}{\includegraphics{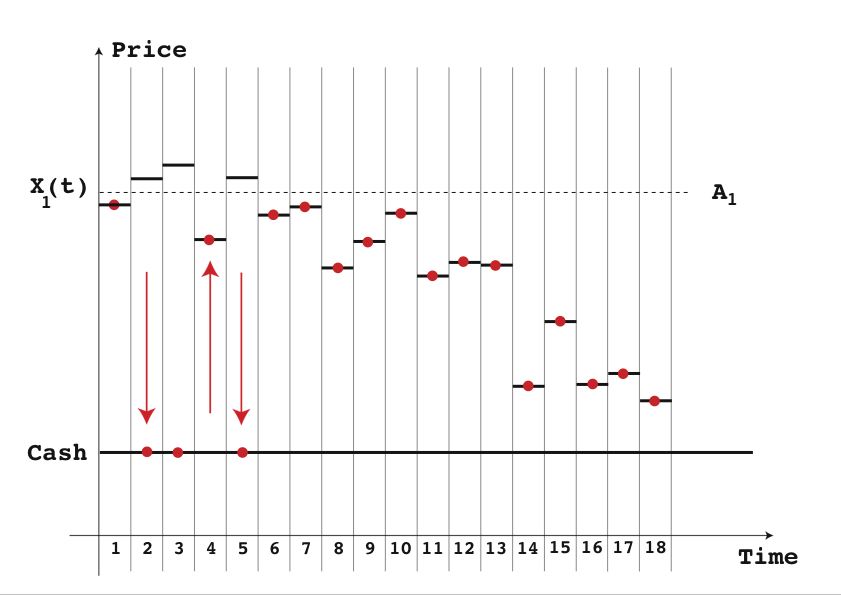}}}
\caption{\label{fig5} The non vanishing number of security does increase with time, however the last security price is so low that the algorithm has lost money at the end of the process ($t_{18}$).}
\end{figure}

Of course, the binomial distribution of the price fluctuations (\ref{basic}) is not a necessary condition for the efficiency of the algorithm. It is just a convenient distribution which allows some easy analytical computations. In fact, provided the knowledge of the future average price, the algorithm can be applied to any more complex $\phi$ distributions, like the ones used in fig.\ref{fig3} or fig.\ref{fig5}, and still leads to an increase of the non vanishing number of security ($N_{1}$) with time (fig.\ref{fig4}).

Now the crucial question is : is something stands in our way and prevent us to use the previous algorithm (MD1) for playing the market?
The answer is yes! First hindrance, the improbable knowledge, at the beginning of the process, of the average security price $A_{1}$. Second possible obstacle, the final price of the security (i.e. the price of the security when one stops to play). Indeed
for the $\phi$ distributions used in fig.\ref{fig1} and fig.\ref{fig3}, the final prices have been arbitrary chosen in a close neighborhood of the average price. But, when it is no more the case as in fig.\ref{fig5}, the final price of the securities may be so low that it can not be balanced by the increase of the non vanishing numbers of securities ($N_{1}$).

In order to face up to these problems, we propose to use, instead of the global average values $A_{1}$, a local one defined thanks to the $m$ last known prices as
\begin{equation}
\left<X_{1}\right>_{m}(t_{n})={{1}\over{m}} \sum_{j=0}^{m-1}X_{1}(t_{n-j})
\label{local}
\end{equation}
such that, at time $t_{n}$, the new demon's behavior (MD2) is ruled by the comparison between $X_{1}(t_{n})$ and $\left<X_{1}\right>_{m}(t_{n})$ in place of $A_{1}$. It is worth noting that, with this modification, we are no more able to prove that the non vanishing numbers of security increase with time, nor that there is gain. Also, the algorithm (MD2) can be optimized by a suitable choice of the $m$ parameter. But, to what is this optimal value related to?

\section{Test with artificial data}
\label{test}
Before testing the algorithm (MD2) with historical data, we first investigate its efficiency with numerous and well controlled artificial aleatory data, satisfying the following modified Geometric Ornstein-Uhlenbeck process \cite{Dixit,Metcalf95}
\begin{equation}
\left\{
\begin{array}{l}
dX_{1}=\alpha_{1}\left(A_{1}-X_{1}\right) dt +\beta_{1} X_{1} dW_{1}
\cr
dA_{1}=\mu_{1} A_{1} dt
\cr
\end{array}
\right.
\label{sode}
\end{equation}
where $W_{1}$ is a Wiener's process. The interpretation of the model is straightforward: when $\alpha_{1}$ is vanishing, eq.(\ref{sode}) describes the well known random walk of the prices. On the contrary, $\alpha_{1}> 0$ means that there exist some fundamentals (liabilities, monopolies, patents or dividends...) which attracts the dynamics of the security price $X_{1}$ in the neighborhood of $A_{1}$. Note that $A_{1}$ is not forced to be constant and may depend exponentially on time if $\mu_{1}\ne 0 $. We proceed in the following way. First we numerically integrate eq(\ref{sode}) over $1000$ units of time using the Milstein's scheme \cite{Milstein78}, a Box-Muller \cite{NumericalRecipies} normal noise generator and $X_{1}(0)=A_{1}(0)$ as an initial condition. Then we apply the modified algorithm (MD2) and compute the associated cumulative return. These last two steps are repeated 30000 times (enough for convergence) and the final result is obtained by averaging.

\begin{figure}[!h]
{\resizebox*{8.5cm}{6.0cm}{\includegraphics{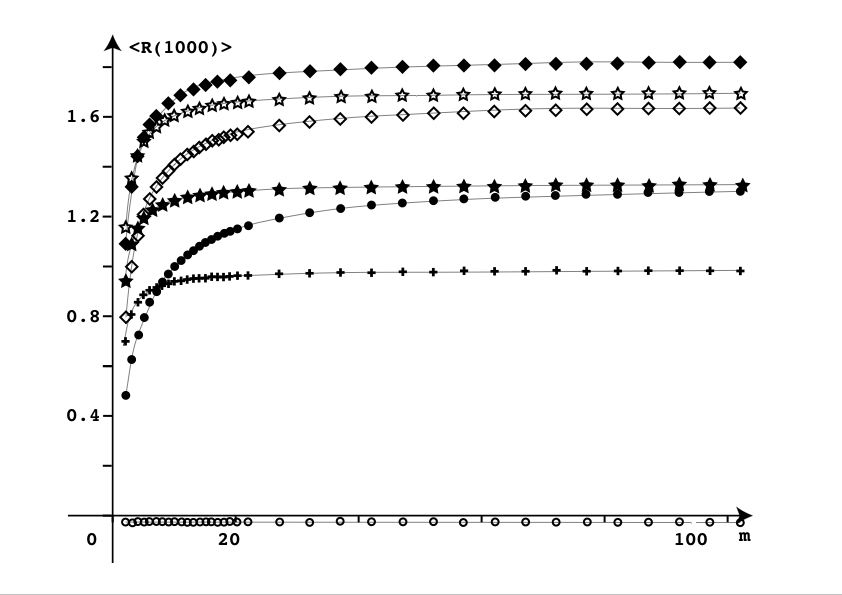}}}
\caption{\label{fig7}Application of the MD2 algorithm onto data generated by (\ref{sode}) with $A_{1}$$=$$2$, $\beta_{1}$$=$$0.01$ and $\mu_{1}$$=$$0$. The plot displays the average cumulative return $\left<R(t_{1000})\right>$ after 1000 units of time versus $m$ for various values of $\alpha_{1}$. From bottom to top, $\alpha_{1}$$=$$0$, $\alpha_{1}$$=$$9.6$, $\alpha_{1}$$=$$0.3$, $\alpha_{1}$$=$$4.8$, $\alpha_{1}$$=$$0.6$, $\alpha_{1}$$=$$2.4$ and $\alpha_{1}$$=$$1.2$.}
\end{figure}

\begin{figure}[!h]
{\resizebox*{8.5cm}{6.0cm}{\includegraphics{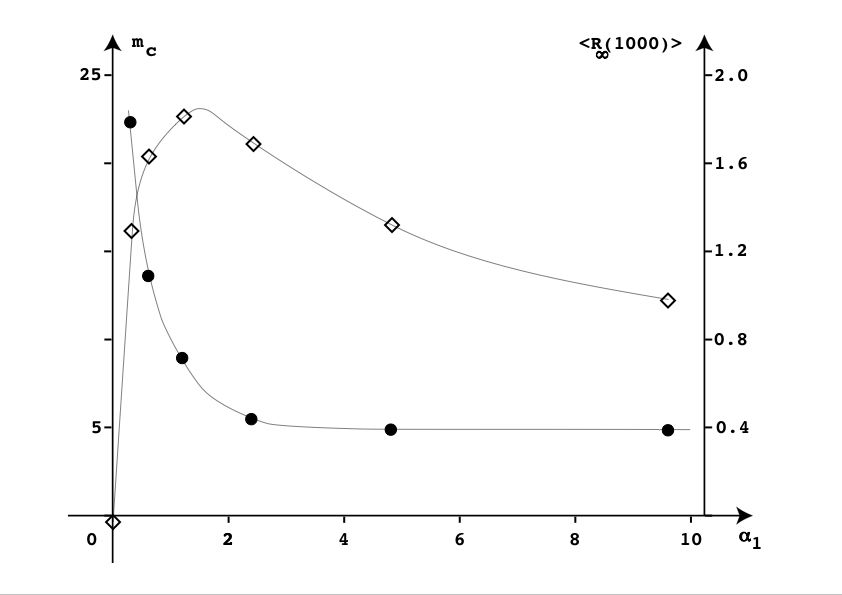}}}
\caption{\label{fig7bis} Same regime of parameters as in fig.\ref{fig7}. The black disks (left scale) correspond to $m_{c}$ versus $\alpha_{1}$ and the white squares (right scale) to the asymptotic values of the average cumulative return $\left< R_{\infty}(t_{1000}) \right>$ versus $\alpha_{1}$. }
\end{figure}

\noindent
In a first step, we restrict ourselves to situations where $A_{1}$ is constant with time. The corresponding results are summarized in fig.(\ref{fig7}) which displays the average cumulative return versus $m$ for various values of $\alpha_{1}$. Several important observations are in order:
\begin{enumerate}
\item In absence of fluctuations (or when $\alpha_{1}>>\beta_{1}$), the cumulative return is just equal to 0 (fig.\ref{fig7bis}). 
\item In presence of fluctuations, but in absence of restoring force (i.e. $\alpha_{1}=0$), the average cumulative return is always vanishing, whatever the value of $m$. 
\item Therefore, a cumulative return distinct from 0 requires the presence of both fluctuations and restoring forces. In such a case, one observe that $\left< R \right>$ increases with $m$ up to an asymptotic value. This increase has to be related to the fact that the more $m$ increases, the more the local average $\left<X_{1}\right>_{m}(t_{n})$ goes close to its asymptotic value $A_{1}$.
\item A characteristic value $m_{c}$ can be defined as the value of $m$ for which the average cumulative return is equal to $90 \%$ of its asymptotic value. As expected, $m_{c}$ decreases with the restoring forces (fig.\ref{fig7bis}).  
\end{enumerate}

\noindent
When $A_{1}$ is no more constant with time, the local average $\left<X_{1}\right>_{m}(t_{n})$ and $A_{1}(t_{n})$ can strongly move away one from the other as $m$ is increased. In such a case, the algorithm (MD2) is certainly less efficient. Hence we expect and do observe (fig.{\ref{fig8}) a decrease of the average cumulative return for sufficiently high value of $m$. The maximum is reached for a value of $m$ close to $m_{c}$ and therefore depends on $\alpha_{1}$. The characteristic time ${{1}\over{\vert \mu_{1} \vert}}$ controls the decrease of the cumulative return after the maximum as well as the amplitude of the maximum.

\begin{figure}[!h]
{\resizebox*{8.5cm}{6.0cm}{\includegraphics{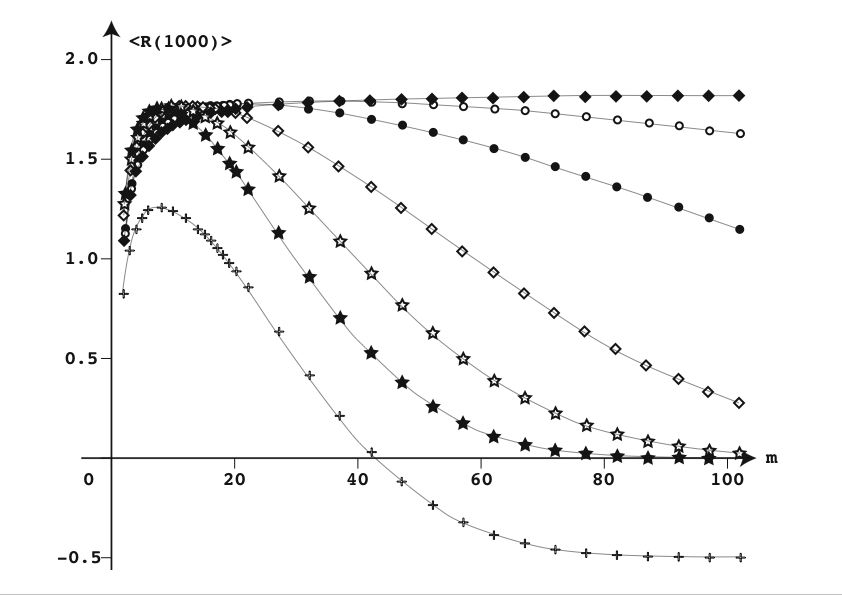}}}
\caption{\label{fig8}Application of the MD2 algorithm onto data generated by (\ref{sode}) with $A_{1}$$=$$2$, $\beta_{1}$$=$$0.01$ and $\alpha_{1}$$=$$1.2$. The plot displays the average cumulative return $\left<R(1000)\right>$ versus $m$ for various values of $\mu_{1}$. From bottom to top, black squares, white disks, black disks, white squares, white stars, black stars and crosses correspond respectively to $\mu_{1}$$=$$0$ $10^{-3}$, $+2$ $10^{-3}$, $+4.0$ $10^{-3}$, $+8.0$ $10^{-3}$, $+12$ $10^{-3}$, $+16$ $10^{-3}$ and $-16$ $10^{-3}$.  }
\end{figure}
Note that the arbitrary chosen exponential dependence of $A_{1}$ with time is not crucial and that similar results would have been obtained with other time dependences (like periodic or even random): only relevant is the existence of two distinct characteristic time scales, one associated with the price evolution, the other, much more slower, associated with the temporal evolution of $A_{1}$.

\section{European Monetary System}
\label{EuMoSy}

We now apply our modified algorithm to genuine data corresponding to the daily record of historical prices. For the sake of efficiency, we look for data for which we expect the existence of a strong restoring force. It turns out that there is no doubt about the existence of such a force for the time evolution of the currencies which belonged to the European Monetary System (EMS). Indeed, the EMS was an economic and politic arrangement where most nations of the European Economic Community linked their currencies to prevent large fluctuations relative to one another. In the early 1990s the European Monetary System was strained by the differing economic policies and conditions of its members, especially the newly reunified Germany, and Britain permanently withdrew from the system (see the 1992 G. Soros's speculation).

In order to map the situation with two currencies onto the previous one, it is sufficient to consider one of the currency as playing the role of the security, and the other as the money in which the value of the security is expressed. The results of the application of the MD2 to the EMS's data are then shown in fig.\ref{fig9}. Obviously, the algorithm works! Even if the transaction costs are taken into account, average yearly returns of $24 \%$ (DEM/GBP) and $44 \%$ (DEM/FRF) are obtained between 1980 and 1992 (13 years), for a suitable choice of $m$. Second and important observation, the optimal value of $m$ does not deeply depend on the choice of the time interval (out of sample). Finally, the curves roughly look like those obtained with (\ref{sode}). Therefore if the theoretical framework (\ref{sode}) holds, we can in addition deduce that: $(\rm{i})$ the restoring forces could be about the same in UK and in France and the optimal value $m_{c} \simeq 5$ could correspond to a weekly action of the central banks. 
$(\rm{ii})$ The characteristic time associated with the decrease of the cumulative return after its maximum is higher for the DEM/GBP than for the DEM/FRF pair. Also the maximum is smaller. Both observations could be related to the well known higher stability of the DEM/FRF's exchange rate (strongly politically motivated by the building of the euro) compared to the DEM/GBP's one (reticence of the UK government).

Fig.(\ref{fig10b}) shows the time evolution of the cumulative daily returns for the application of the MD2 with $m=5$. The main features of the curves are the quite regular growth between 1980 and 1990 and its abrupt stop in the early 1990s. In the framework of (\ref{sode}), this stop is understood as the vanishing of the restoring forces. Note the stop date is in perfect agreement with the acknowledged beginning of the EMS's trouble. 
\begin{figure}[!h]
{\resizebox*{8.6cm}{6.0cm}{\includegraphics{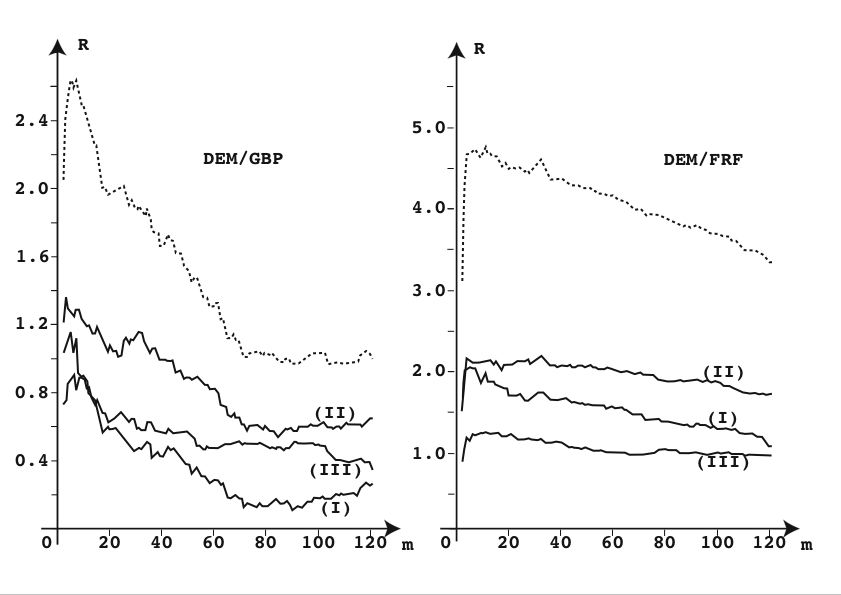}}}
\caption{\label{fig9} Application of the MD2 to the historical EMS currencies. The plots display the cumulative return $R$ versus $m$. A cost of $3/10000$ per transaction has been applied. The dashed lines correspond to the 1980-1992 time interval. For the continuous lines, the time interval 1980-1990 has been splitted into 3 equal parts (quoted {\bf I}, {\bf II} and {\bf III}).}
\end{figure}

\begin{figure}[!h]
{\resizebox*{8.6cm}{4.0cm}{\includegraphics{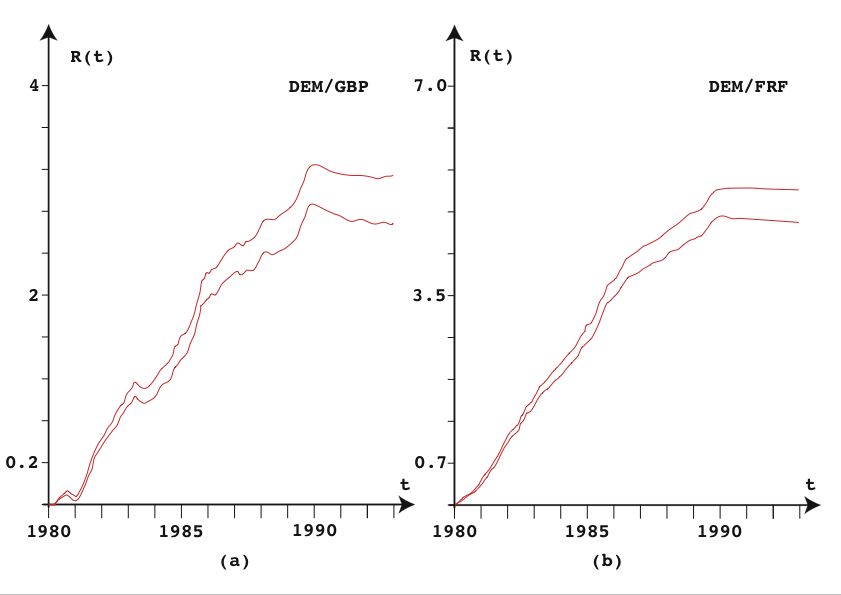}}}
\caption{\label{fig10b} Application of the MD2 with $m$$=$$5$ to the  EMS currencies. The plots display the cumulative daily return versus time. The top curves correspond to a free cost transaction while for the lower one a cost of $3/10000$ has been applied.}
\end{figure}

\section{Stock Exchange}
\label{stock}
The next question is: what happens with the stock exchange? In one hand, one can expect the existence of liabilities, patents, company's reputation and monopoly to lead to the existence of a restoring forces. But in the other hand, rumors, non informed agents, mimicry and speculation play an opposite role and can 
drive the security prices far from their fundamental values. In the following, we investigate the efficiency of our algorithm onto the components of the Dow Jones, Dax and Cac40 indexes, between 2000 and 2006. We proceed in the following way. Among the components of each index, the securities which are not regularly quoted for the time interval between 2000/01/01 and 2006/05/12  ($\simeq 6.5$ years) are thrown away. We are then left with $N_{s}$$=$$30$ securities for the Dow Jones, $N_{s}$$=$$29$ for the Cac40 and $N_{s}$$=$$24$ for the Dax. Then, for each index, we restrict our analysis to the days for which all the securities which remain after the first selection are quoted. We then apply MD2 separately to each component $j$ of a single index. "Separately" means that the money which is obtained with a given security is not used to speculate with an other one. The same initial investment ${\cal W}(0)$ is used for each component. At time $t$, the cumulative return of a given index is then defined as
\begin{equation}
R_{index}=log\left( {{\sum_{j=1}^{N_{s}}{\cal W}_{j}(t)}\over{N_{s}{\cal W}(0)}} \right)
\label{indexReturn}
\end{equation}
where ${\cal W}_{j}(t)$ stands for the wealth which is obtained playing MD2 with the  security $j$. Transaction costs of $0.1 \%$ have been applied. The corresponding results are then displayed in fig.\ref{fig13} for the Cac40, fig.\ref{fig14} for the Dax and fig.\ref{fig15} for the Dow Jones.

\begin{figure}[!h]
{\resizebox*{8.6cm}{4.0cm}{\includegraphics{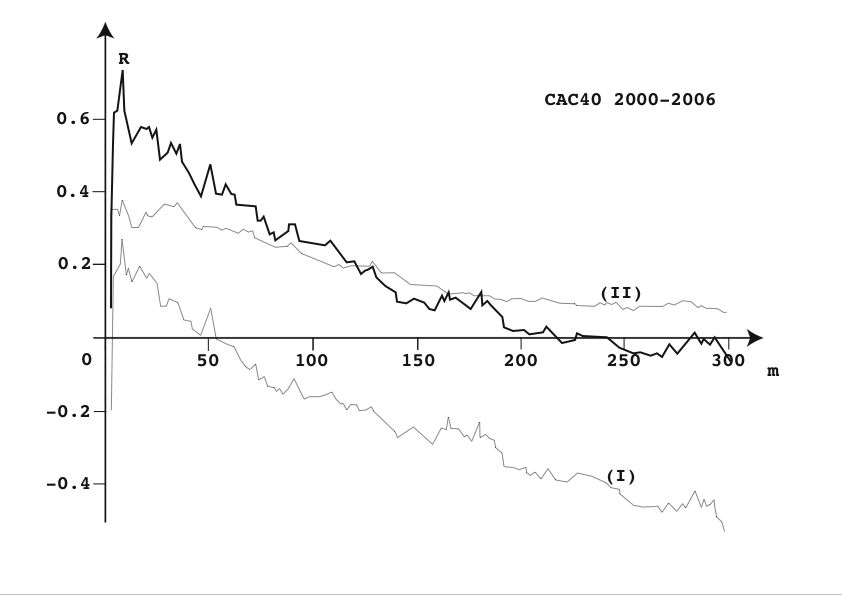}}}
\caption{\label{fig13} Application of MD2 algorithm to the Cac40 index. The 3 continuous lines correspond to the plot of the cumulative return (\ref{indexReturn}) versus $m$, but the thick one deals with the historical records between 2000 and 2006, while the thin ones are associated respectively with the first (I) and second (II) half of the same time interval.}
\end{figure}

\begin{figure}[!h]
{\resizebox*{8.6cm}{4.0cm}{\includegraphics{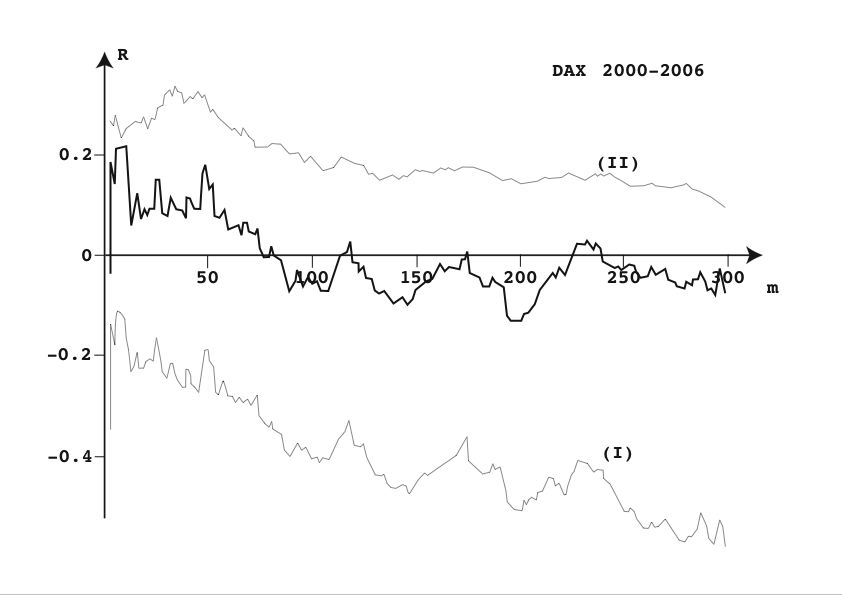}}}
\caption{\label{fig14} Same caption as fig.\ref{fig13} but for the Dax index.}
\end{figure}

\begin{figure}[!h]
{\resizebox*{8.6cm}{4.0cm}{\includegraphics{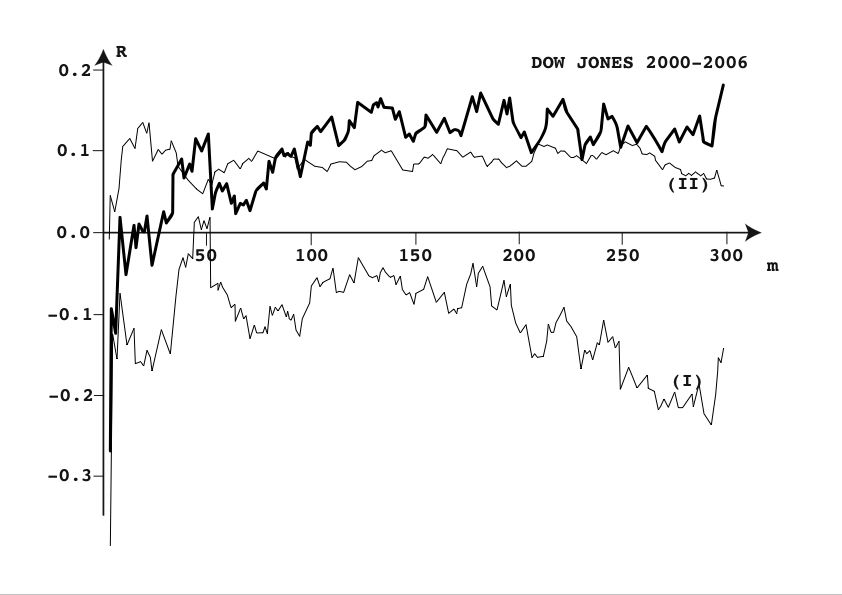}}}
\caption{\label{fig15} Same caption as fig.\ref{fig13} but for the Dow Jones index.}
\end{figure}

Several remarks are in order:
$(\rm{i})$ for each index, the shape of the curves associated with either the wholeness or halves of the time interval, looks like the same, $(\rm{ii})$ the european and US behaviors are deeply different. Roughly, for large values of $m$, we observe a decrease of the cumulative return for the Cac40 and Dax, and an increase for the Dow Jones. In the framework of (\ref{sode}), this observation could suggest the existence of a stronger restoring force for the european indexes than for the US one, $(\rm{iii})$ for the whole time interval, the best european results are obtained for $m \simeq 8$. For the Cac40, the maximum value is close to $0.7$ and corresponds to a yearly average return of $\simeq 11 \%$. For the Dax, the yearly return is smaller and close to $\simeq 3.5 \%$. The smallest yearly return ($\simeq 2\%$) is obtained for the Dow Jones. Although small, the yearly returns provided by the MD2 algorithm over the whole time interval are always better than the buy and hold strategy. Indeed, this last strategy leads, for the same time interval and the same selected securities, to an average yearly return of $\simeq 0.0\%$ for the Cac40, $\simeq -0.5\%$ for both the Dax and the Dow Jones.

\section{The fabulous case of the Cac40}
\label{fabulous}

Concerning the financial market crashes, the french exception has already been brought to light \cite{SornetteCrash}. As an other evidence of this strangeness, we now discuss a modification of the MD2 algorithm which will reveal itself as especially competitive for the Cac40 market, but of mediocre interest for Dax and without interest for the Dow Jones. 

In the previous algorithm (MD2), each security was considered independently from the other because the amount of money which is assigned to a particular security can not be used to buy an other security (eq. \ref{indexReturn}). This constraint is now relaxed in the new algorithm (MD3), which is defined as:
\begin{enumerate}
\item The daemon is playing once at each time $t_{n}$.
\item At a given time, the daemon is in possession of only one type of security, cash being considered as a special type of security.
\item At time $t_{n}$, the choice of the new position is given by the following rule: Assume that there are $N_{s}$ securities whose prices are $X_{1}(t)$, $X_{2}(t)$,....$X_{N_{s}}(t)$. Assume also that at $t=t_{n}$, the Maxwell daemon is in possession of $N_{j}(t_{n})$ securities of type $j$ ($k \ne j \Longrightarrow N_{k}=0$). Defining $r_{j,k}$ and $\sigma_{j,k}$ as
\begin{equation}
\begin{array}{l}
r_{j,k}={{1}\over{m}}\sum_{p=0}^{m-1}{{X_{j}(t_{n-p})}\over{X_{k}(t_{n-p})}}
\cr
\sigma_{j,k}=\sqrt{{{1}\over{m}}\sum_{p=0}^{m-1}\left({{X_{j}(t_{n-p})}\over{X_{k}(t_{n-p})}}\right)^2-r_{j,k}^2}
\end{array}
\label{md3Definitions}
\end{equation}
then the new position correspond to the $k$ value which maximizes
\begin{equation}
{{{{X_{j}(t_{n})}\over{X_{k}(t_{n})}}-r_{j,k}}\over{\sigma_{j,k}}}
\label{md3Optimized}
\end{equation}
\item As for MD2, $m$ is a free parameter which stands for the number of past data used to compute the mobile averages.
\item The daemon starts with cash (${\cal W}(0)$) and the cumulative return at time $t_{n}$ is defined as $ln\left({{{\cal W}(t_{n})}\over{{\cal W}(0)}}\right)$.
\end{enumerate}
Fig.(\ref{fig16}) displays the stupendous results of the application of the MD3 algorithm to the components of the Cac40 between 2000/01/01 and 2006/05/12  ($\simeq 6.5$ years). First, the optimal value of $m$ does not depend on the time interval (out of sample). Second this optimal value is found to be close to 25 days, so to say one month since only workdays are taken into account. Finally but not the least, average yeraly return up tu $60\%$ are obtained!

\begin{figure}[!h]
{\resizebox*{8.6cm}{4.0cm}{\includegraphics{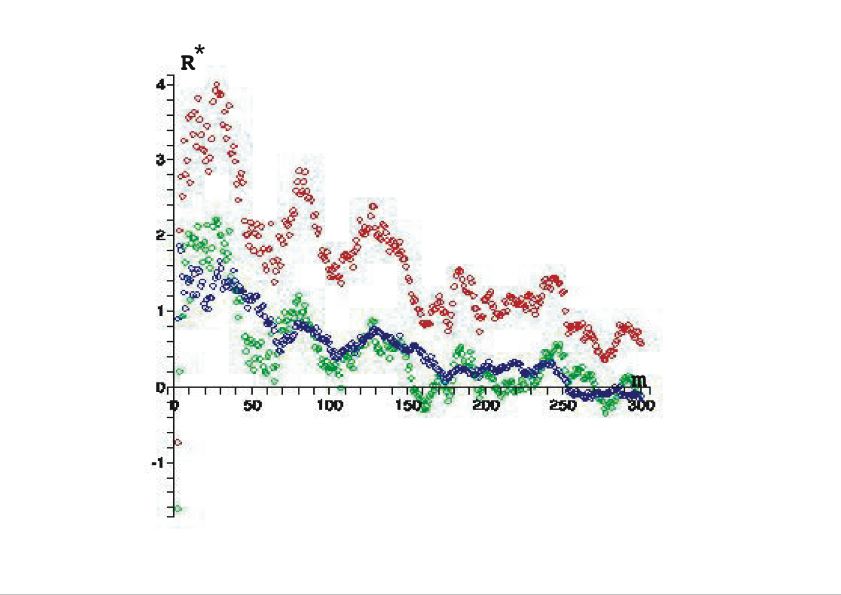}}}
\caption{\label{fig16} Application of the MD3 algorithm to the Cac40 index. $0.1\%$ transaction cost have been applied. The picture display the cumulative return versus $m$. The red points are associated with the cumulatice return between 2000 and 2006, while the green and the blue ones are associated respectively with the first (I) and second (II) half of the same time interval. }
\end{figure}

\section{Conclusion}
\label{conc}
By analogy with the way motor enzymes trap favourable brownian fluctuations, we have built an algorithm which is able to 
make the best from out of equilibrium price fluctuations and to play the market. Testing its efficiency with genuine historical data, positive cumulative returns have been measured even in presence of a $0.1\%$ transaction cost. Especially stupendous are the results dealing with the application of the algorithm to EMS currencies or with the Cac40 components.

Several remarks are in order:
\begin{enumerate}
\item With no doubt, the birth of the EMS was first a strong political resoluteness. However, numerous economics experts, governors of the central banks and members of the european monetary comity have been lengthily consulted. How is it possible that such an opportunity to make money at the expense of nations has not been identified?
\item The money which is captured by our algorithms comes from the irrational behavior of uninformed noisy traders. Therefore we really expect the present algorithms will become unprofitable as soon as our paper will be published, either because irrational traders will be taught a lesson or because the profitability of the algorithms will vanish with the number of users.
\item More than a way to make money, the algorithms can be used as simple, sensitive and straightforward tools to detect non efficient market, certainly less circuitous than the measure of the excess volatility or the deviation from the normal distribution.
\end{enumerate}

\vskip 0.5 true cm
\noindent
The author is grateful to J. Viting-Andersen and V. Planas-Bielsa for valuable discussions.

\end{document}